\documentclass[runningheads]{llncs}
\usepackage{graphicx}
\usepackage{amsmath}
\usepackage{amssymb}
\usepackage{environ}
\usepackage{floatrow}
\usepackage{tabularx}
\usepackage{hyperref}
\usepackage[utf8]{inputenc}
\usepackage[T1]{fontenc}
\PassOptionsToPackage{hyphens}{url}\usepackage{hyperref}
\usepackage{graphicx,multirow}
\usepackage{array}
\setbox0\hbox{\tabular{@{}l}XML\endtabular}
\setbox1\hbox{\tabular{@{}l}SQL\endtabular}
\begin{document}
\title{Novel version of PageRank, CheiRank and 2DRank for Wikipedia in Multilingual Network using Social Impact}
\titlerunning{Novel version of PageRank, CheiRank and 2DRank...}
%
%
\author{Célestin Coquidé\inst{1}\orcidID{0000-0001-8546-6587} and
Włodzimierz Lewoniewski\inst{2}\orcidID{0000-0002-0163-5492}}
\authorrunning{C. Coquidé and W. Lewoniewski}
%
\institute{Institut UTINAM, Observatoire des Sciences de l'Univers THETA, CNRS,
Universit\'e de Bourgogne Franche-Comt\'e, Besançon, France.
\email{celestin.coquide@utinam.cnrs.fr}\\
\and
Poznan University of Economics, Department of Information Systems, Poznan, Poland.\\
\email{Wlodzimierz.Lewoniewski@ue.poznan.pl}}
\maketitle              
\begin{abstract}
Nowadays, information describing navigation behaviour of internet users are used in several fields, e-commerce, economy, sociology and data science. Such information can be extracted from different knowledge bases, including business-oriented ones. In this paper, we propose a new model for the PageRank, CheiRank and 2DRank algorithm based on the use of clickstream and pageviews data in the google matrix construction. We used data from Wikipedia and analysed links between over 20 million articles from 11 language editions. We extracted over 1.4 billion source-destination pairs of articles from SQL dumps and more than 700 million pairs from XML dumps. Additionally, we unified the pairs based on the analysis of redirect pages and removed all duplicates. Moreover, we also created a bigger network of Wikipedia articles based on all considered language versions and obtained multilingual measures. Based on real data, we discussed the difference between standard PageRank, Cheirank, 2DRank and measures obtained based on our approach in separate languages and multilingual network of Wikipedia.
\end{abstract}
%
%
%
\section{Introduction}
For the last 10 years, Wikipedia has been one of the most popular source of knowledge. In different countries, this online encyclopedia is in the top 10 most visited websites \cite{url101}. Wikipedia  even influences the language used by scientists in their publications~\cite{Thompson2018}. Nowadays, this free encyclopedia contains over 51 million articles in more than 300 languages \cite{url102}. Content on a separate subject can be created and edited independently in each language version of Wikipedia \cite{lewoniewski2019comp}. Often a Wikipedia article contains links to other pages, which can be used to find more information related to the subject. Based on links between articles, it is possible to identify important places, persons, products in specific area or language community. To find such important Wikipedia articles, different methods can be used. One of the well-known algorithms for this purpose is PageRank \cite{page1999}.

The google matrix and the PageRank algorithm are at the foundation of the very famous search browser Google \cite{brin98}. They describe a random internet user, surfing on the World Wide Web (WWW). Indeed WWW can be seen as a directed network where nodes are Web pages and links are hyperlinks allowing an internet user to navigate. By knowing a network's topology we can construct its adjacency matrix $A$ as well as its stochastic matrix $S$. Its mathematical aspect is well described in \cite{meyer_2012}. Centrality measures in complex network theory are important, especially as nowadays networks are very large and complex (high occurrence of nodes and links). In case of directed networks such as WWW, eigenvector-based centrality measure is helpful and efficient as demonstrated by Google's web search engine. However, this simple random walk model may be limited for describing a human user's behaviour. PageRank algorithm, applied to the Wikipedia network, leads to robust ranking lists of articles and articles of countries tend to be well ranked. Using a simple random walk, such algorithm didn't allow us to observe a trend of interest from readers.

In this study, we extend the usual PageRank algorithm with clickstream's and pageview's data from Wikipedia October 2019's dumps. In this paper, we first mention works related to Wikipedia and PageRank algorithm, then we present you the ''WikiClick'' ({\textbf{\textit{wc}}}) and ''WikiClick Plus View'' ({\textbf{\textit{wcpv}}}) models and finally we discuss how this social information affects Wikipedia articles ranking.

\section{Related works}
Pageviews statistics in Wikipedia \cite{url8} can show which article is popular over a specific period. Therefore, we can observe a topic trend over time. Using such data, it is possible to forecast stock market moves \cite{moat2013}, movies' success \cite{mestyan2013,latif2016}, demand for services in the tourism sector \cite{khadivi2016}, cryptocurrencies price and market performance \cite{kristoufek2013,elbahrawy2019}, epidemics in specified territory\cite{hickmann2015} and can be used in electoral prediction \cite{yasseri2016}. Moreover, it can be used to assess the quality of Wikipedia articles alongside other measures (such as text length, number of references, images, sections and others)\cite{lewoniewski2019,lewoniewski2019comp}.

Google matrix and PageRank algorithm have been applied to Wikipedia for ranking articles related to historical figures \cite{eom15} and Universities \cite{lages16}. We can also find studies on the World influence for Universities \cite{coquide19-1}, cancers \cite{rollin19-1} and infectious diseases \cite{rollin19-2} using the Wikipedia network. The Wikipedia articles ranking evolution over time using PageRank and CheiRank for countries, historical figures, physicist and chess players  have been investigated in \cite{eom-time-13}. Spectral analysis of the Google matrix permits us to retrieve communities of articles \cite{ermann13}. Moreover the google matrix analysis have been used in business oriented networks such as the World trade network \cite{coquid2020crisis} and cryptocurrencies network \cite{coquid2019bit}.

The use of the internet users' behaviour information such as pageviews and clickstream data has been studied for the biomedical ontology repository BioPortal \cite{espin19}. Pageviews data is used by pantheon group to rank famous individuals from Wikipedia \cite{yu16}. Clickstream data has been used to study the navigational phase space of Wikipedia \cite{gildersleve18}.

\section{Proposed methods}
\subsection{Wikinetwork models}
In such network, nodes are Wikipedia articles and the directed link $A\rightarrow B$ exists if article $B$ is reachable from article $A$ by using an intra-wiki hyperlink.
Usually, we consider the unity rule for such links, it means that we count once such hyperlink in article $A$. It follows that the corresponding adjacency matrix will be asymmetric and its elements are equal to $1$ or $0$. We call this model the standard wikinetwork model ({\textbf{\textit{nowc}}}).

Information from wikidata such as clickstream for intra-wiki hyperlinks provides an interesting social bias of the Wikipedia network. We construct two different models of Wikipedia network: wikiclick ({\textbf{\textit{wc}}}) and wikiclick plus view ({\textbf{\textit{wcpv}}}). We compare results obtained using these methods with {\textbf{\textit{nowc}}}. 

\subsection{Google matrix and Ranking algorithms}
\subsubsection{Google matrix}

The google matrix $G$ is a Perron-Frobenius operator based on the stochastic matrix of a network and a teleportation term. We detail here the construction of $G$ for each model.

The general formula for the google matrix is
\begin{equation}
G=\alpha S + (1- \alpha) \mathbf{v} \mathbf{e^{T}}
\label{Gequation}
\end{equation}

where $\alpha$ is the so-called damping factor, $S$ the stochastic matrix, $\mathbf{e^{T}}$ a row vector with $N$ ones and $\mathbf{v}$ is a preferential vector such as $\sum_{j}v_{j} = 1$. $\alpha = 0.85$ is a standard value for different real complex networks, such as WWW.

The teleportation term $\mathbf{v} \mathbf{e^{T}}$ may simply be a matrix of elements equal to $1/N$.

The google matrix for the standard model is computed with this trivial teleportation term.

$S$ is computed from the adjacency matrix describing the network topology
\begin{equation}
A_{ij}=
\left\{\begin{array}{ll}   
1& {\text{ if edge }}j\rightarrow i{\text{ exists }}\\
0&{\text{ else }}
\end{array}\right.
\label{eqA0}
\end{equation}

In case of {\textbf{\textit{wc}}} version of Wikipedia network, we simply use $W$ the matrix of clicks where $W_{ij}$ element is the number of clicks received, article $i$ from article $j$.
The Wikidata for clickstream only counts clicks $\geq 10$, $W_{ij} = 0$, therefore elements are replaced by a standard $A_{ij}$ element representing the possibility of click because the link exist in the network, this final weighted adjacency matrix is noted $A_{wc}$.
From $A_{wc}$ we define the stochastic matrix $S_{wc}$ representing the probability to reach node $i$ from $j$ by:

\begin{equation}
S_{wc_{ij}} = \left\{\begin{array}{ll}
\frac{A_{wc_{ij}}}{\sum_{i'}A_{wc_{i'j}}}&\text{ if $\sum_{i'}A_{wc_{i'j}} \neq 0$ }\\
1/N &\text{ if $\sum_{i'}A_{wc_{i'j}} = 0$ }
\end{array}
\right.
\label{eqS}
\end{equation}

We also use pageviews information as a teleportation matrix instead of $\mathbf{v} \mathbf{e^{T}}$. In that way, the preferential vector component $v_{j} = \#$views for article $j$. $\mathbf{\Tilde{v}}$ is the normalized vector computed from $\mathbf{v}$ such as $\sum_{j}\Tilde{v_{j}} = 1$.

Finally, we obtain:
\begin{equation}
    G_{ij} = \left\{\begin{array}{ll}
    \alpha S_{wc_{ij}} + (1-\alpha)/N &{\text{ for {\textbf{\textit{wc}}}}}\\
    \alpha S_{wc_{ij}} + (1-\alpha)\Tilde{v_{i}}&{\text{ for {\textbf{\textit{wcpv}}}}}
    \end{array}
    \right.
    \label{eqGbis}
\end{equation}
\subsubsection{PageRank and CheiRank algorithm}
The leading right eigenvector of $G$ with corresponding eigenvalue $\lambda = 1$ corresponds to the steady state of a random walker moving through the network for an infinite time.
We have the relation
\begin{equation}
    G \textbf{P} = \textbf{P}
    \label{preq}
\end{equation}
where $\mathbf{P}$ is called PageRank vector, its $i^{\text{th}}$ component represents the probability that a random surfer reaches node $i$ after an infinite journey.
By sorting components in decreasing order, we obtain the nodes ranking. PageRank measures ingoing links efficiency, seen as importance ranking.
Let $K_{1}, K_{2},..., K_{N}$ be the PageRanks of the nodes such as $P_{K_{1}} > P_{K_{2}}>P_{K_{3}}$ and so on.

To measure the efficiency of outgoing links, we simply reverse the direction of all the network's links. This leads to a new adjacency matrix $A_{wc}^{*} = A_ {wc}^{T}$ and then we compute the corresponding google matrix noted $G^{*}$.

CheiRank vector ($\mathbf{P^{*}}$) is defined as the eigenvector of $G^{*}$ such as $G^{*} \mathbf{P^{*}} = \mathbf{P^{*}}$ and note $K^{*}_{1}, K^{*}_{2},..., K^{*}_{N}$ the obtained ranking of nodes.
\subsubsection{2DRank algorithm}
PageRank and CheiRank algorithms are two sides of the same coin, the first one describes relevance of an article within Wikipedia's network and the last one represents its communicability. As described in \cite{zhirov10,ermann12}, we can place nodes of a network in the two-dimensional $(K,K^{*})$ plane. The 2DRank algorithm uses both rankings and is defined as following:
\begin{itemize}
    \item Firstly, for each node let, $K_{max}(i) = Max(K_{i}, K^{*}_{i})$.
    \item Secondly, we sort nodes in ascending order according to their $K_{max}$.
    \item Finally, we sort couple of nodes with the same $K_{max}$ regarding the increasing $K^{*}$ ordering.
\end{itemize}

\section{Datasets and extraction methods}
To extract data about links between Wikipedia articles (wikilinks or intra-wiki), we use Wikipedia dumps from October of 2019 \cite{url1}. We focused on two separate approaches to identify these links:
\begin{itemize}
    \item \textbf{XML} - directly from the Wikicode \cite{url2} (XML dumps)
    \item \textbf{SQL} - rendered versions of the articles (SQL dumps).
\end{itemize}  

In the case of Wikicode, for each article, we searched internal links (wikilinks) \cite{url3} placed in doubled square brackets in code for each considered language version (below example for English Wikipedia):
\begin{itemize}
    \item ''enwiki-20191001-pages-articles.xml.bz2'' - recombined articles, templates, media/file
    descriptions, and primary meta-pages.
    \item ''enwiki-20191001-redirect.sql.gz'' - redirect list.
\end{itemize}
Among the extracted links were also the ones that led to other types of Wikipedia pages (other namespaces \cite{url4}). Therefore, we only kept those belonging to article namespace (ns 0). We also removed links leading  to nonexistent articles (so called ''red links'' \cite{url5}). Additionally, we took into account other names of the same articles based on redirects \cite{url6}.

Rendered version of the articles usually have more links to other pages than we can find in their source (in Wikicode). It comes from additional elements placed in the article. For example, we can find the same template with certain set of links in articles related to similar topic (such as French cities, cryptocurrencies, processors, Nobel Prize laureates and others). In order to analyze links in rendered version of the Wikipedia articles, we took into account other dumps files for each language version (below is an example for the English Wikipedia): 
\begin{itemize}
    \item ''enwiki-20191001-pagelinks.sql.gz'' - wiki page-to-page link records.
    \item ''enwiki-20191001-page.sql.gz'' - base per-page data (id, title, old restrictions, etc).
    \item ''enwiki-20191001-redirect.sql.gz'' - redirect list.
\end{itemize}

The table \ref{tab:pairscount} shows statistics of the extraction source-destination pairs of links for each considered language version of Wikipedia. We extracted over 1 billion pairs from the SQL dump and over 500 million pairs from XML dumps. For every pair, we only took Wikipedia articles from ns 0. After excluding duplicate pairs, redlinks (nonexistent pages) and unification of the articles names (based on redirects), the total pair number is reduced.

\begin{table}[]
\caption{All and unified source-destination pairs of links from Wikipedia in different languages. Source: own calculations based on Wikimedia dumps in October 2019}
\setlength{\tabcolsep}{6pt}
\begin{scriptsize}
\begin{tabular}{p{1cm} | r r | r r}
\hline
\multirow{2}{*}{\textbf{Lang.}} & \multicolumn{2}{|c}{\textbf{XML}}                                        & \multicolumn{2}{|c}{\textbf{SQL}}                                        \\
                                   & \multicolumn{1}{|c}{\textbf{All}} & \multicolumn{1}{c}{\textbf{Unified}} & \multicolumn{1}{|c}{\textbf{All}} & \multicolumn{1}{c}{\textbf{Unified}} \\
                                   \hline
de                                 & 86 242 247                       & 63 618 326                           & 111 288 696                      & 108 762 081                          \\
en                                 & 228 373 266                      & 165 832 345                          & 500 144 739                      & 479 163 241                          \\
es                                 & 54 878 393                       & 39 369 961                           & 53 948 827                       & 50 625 623                           \\
fa                                 & 15 298 097                       & 7 427 045                            & 74 249 078                       & 71 867 560                           \\
fr                                 & 80 719 270                       & 61 576 083                           & 156 691 399                      & 153 108 004                          \\
it                                 & 52 731 642                       & 40 857 564                           & 117 826 190                      & 115 641 441                          \\
ja                                 & 63 112 674                       & 50 122 887                           & 92 626 350                       & 90 901 975                           \\
pl                                 & 36 838 878                       & 27 240 200                           & 76 958 914                       & 76 318 086                           \\
pt                                 & 31 311 443                       & 22 167 152                           & 61 269 416                       & 58 843 986                           \\
ru                                 & 52 646 408                       & 37 922 206                           & 99 995 034                       & 95 706 281                           \\
zh                                 & 30 253 747                       & 18 718 463                           & 86 343 098                       & 83 272 015                           \\
        
\end{tabular}
\end{scriptsize}
\label{tab:pairscount}
\end{table}

For both methods we additionally extracted pageviews statistics \cite{url8}  and clickstream data \cite{url7} (click counts of source-destination pairs of articles) from September 2019.

\section{Application of Methods and Discussion}
In this section, we detail the application of our method using different dumps of Wikipedia from October 2019. We have built the network from English edition of Wikipedia from XML and SQL dumps as well as a multilingual version of the network. In the multilingual network, we took all language editions with available clickstream and pageviews data. 

\subsection{English edition}
\textbf{XML dumps}. Table~\ref{tab:top20prenwiki19} shows the top 10 articles using \textbf{\textit{wcpv}} method and PageRank algorithm. For each of these articles, we also show its rank among $K_{wc}$, $K_{nowc}$, clickstream ($K_{cR}$) and pageviews ranking lists ($K_{vR}$). 

We found articles related to sovereign states for \textbf{\textit{wcpv}} method, which is usually the case for \textbf{\textit{nowc}} PageRank. By comparing the top 10 from $K_{wcpv}$ with $K_{nowc}$ we see that this set of articles is a reordering of $K_{nowc}$ except for "Wikipedia", "List of Queen of the South episodes" and ''Queen of the South (TV series)''. These three articles are badly ranked in $K_{nowc}$ as well as in $K_{wc}$. The two first elements are well ranked because of their Pageviews. The third one is very interesting because ''Queen of the South (TV series)'' is badly ranked in all other ranking list. According to our method, we found a top 10 PageRank containing two articles related to the same TV series. This last result is not common for Wikipedia PageRank.
\begin{table}[]
\caption{First 10 articles obtained by PageRank with {\textbf{\textit{wcpv}}} model from English Wikipedia. Source: own calculations based on Wikimedia dumps in October 2019.}

\centering
\resizebox{\columnwidth}{!}{
\begin{scriptsize}
\begin{tabular}{|p{6cm}|c|c|c|c|c|}
\hline
Name&$K_{wcpv}$&$K_{wc}$&$K_{nowc}$&$K_{cR}$&$K_{vR}$\\
\hline
United States&1&1&1&15&24\\
\hline
Wikipedia&2&11665&3542&25013&1\\
\hline
List of Queen of the South episodes&3&5170889&5128933&4455336&2\\
\hline
United Kingdom&4&9&5&81&63\\
\hline
New York City&5&23&10&150&139\\
\hline
World War II&6&12&3&181&78\\
\hline
Germany&7&7&7&727&118\\
\hline
India&8&10&9&138&68\\
\hline
France&9&5&2&1432&197\\
\hline
Queen of the South (TV series)&10&166342&744237&28297&5871\\
\hline
\end{tabular}
\end{scriptsize}
}
\label{tab:top20prenwiki19}

\end{table}

The top 10 articles from CheiRank according to {\textbf{\textit{wcpv}}} method is detailed in table~ \ref{tab:top20crenwiki19}.
Usually, using the CheiRank method applied to Wikipedia network, articles related to a list of articles have a better rank than others. We do not see that in {\textbf{\textit{wcpv}}}. Indeed, we only have 4 lists (''Deaths in 2019, Lists of deaths by year'', ''2019 in film'' and ''List of Bollywood films of 2019'') whereas \textbf{\textit{nowc}} shows 100$\%$ of lists in its top 10. In the other articles of top 10 \textbf{\textit{wcpv}} CheiRank, some are related to social interest such as ''Joker (2019 film)'' and "2019 FIBA Basketball World Cup". \textbf{\textit{wcpv}} CheiRank is similar to {\textbf{\textit{wc}}}. We think that with the use of clickstream and pageviews, CheiRank algorithm gives us a list of entry points of Wikipedia.

\begin{table}[]
\caption{First 10 articles obtained by CheiRank with {\textbf{\textit{wcpv}}} model from English Wikipedia. Source: own calculations based on Wikimedia dumps in October 2019.}
\centering
\resizebox{\columnwidth}{!}{
\begin{scriptsize}
\begin{tabular}{|p{6cm}|c|c|c|c|c|}
\hline
Name&$K^{*}_{wcpv}$&$K^{*}_{wc}$&$K^{*}_{nowc}$&$K_{cR}$&$K_{vR}$\\
\hline
Deaths in 2019&1&2&909&406&4\\
\hline
Lists of deaths by year&2&1&10&55939&7874\\
\hline
It Chapter Two&3&10&334575&2&12\\
\hline
2019 in film&4&9&382&365&34\\
\hline
List of Bollywood films of 2019&5&12&19249&640&54\\
\hline
Wikipedia&6&421&11031&25013&1\\
\hline
It (2017 film)&7&19&106231&18&37\\
\hline
Joker (2019 film)&8&20&95145&33&10\\
\hline
Mindhunter (TV series)&9&17&310462&147&35\\
\hline
2019 FIBA Basketball World Cup&10&11&20498&44&13\\
\hline
\end{tabular}
\end{scriptsize}
}
\label{tab:top20crenwiki19}

\end{table}

\begin{figure}
\includegraphics[width=0.9\textwidth]{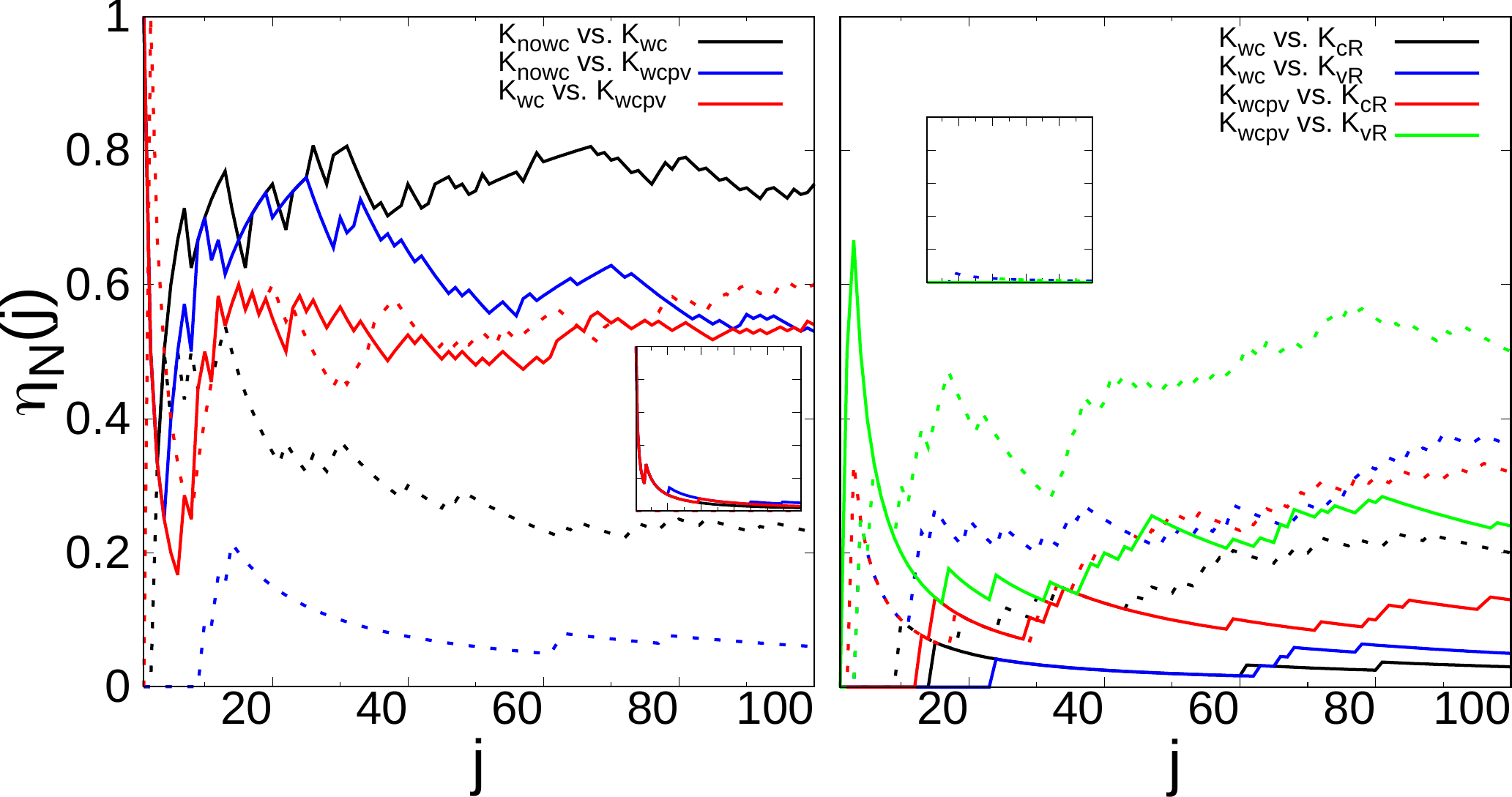}
\caption{Overlap $\eta_{N}$ versus rank $j$ for doublet of ranking lists computed from \textbf{\textit{wc}}, \textbf{\textit{wcpv}} and \textbf{\textit{nowc}} models (left panels),  from {\textbf{\textit{wc}}}, {\textbf{\textit{wcpv}}}, \textit{cR} and \textit{vR} (right panels) considering English edition. Inset plots correspond to exact overlap $\eta_{O}$. Solid and dotted lines are for PageRanks and CheiRanks.} \label{fig1}
\end{figure}

In order to quantify changes in ranking of Wikipedia articles, coming from the used model, we computed two overlap measures $\eta_{N}$ and $\eta_{O}$. The first one measures the presence of same articles in two ranking lists and $\eta_{O}$ measures the exact rank similarity. Quantitatively, by regarding fig.~ \ref{fig1} differences between rankings are related to rank switching  rather than exact similarity. As we can see $\eta_{N}$ is higher than $\eta_{O}$ (inset plots). Moreover, at short range $j \in [1,20]$, we have the highest overlap measures, with $\eta_{N} = 0.75 (K_{nowc}\text{ vs. }K_{wc}), 0.7 (K_{nowc} \text{ vs. }K_{wcpv}$ for $j = 20$. In case of CheiRank overlap, we respectively have $\eta_{N} = 0.35$ and $\eta_{N} = 0.15$ comparing ${K^{*}}_{wc}$ and ${K^{*}}_{wcpv}$ with ${K^{*}}_{nowc}$. Social information seems to change more drastically CheiRank than PageRank. When comparing \textbf{\textit{wcpv}} and \textbf{\textit{wc}} methods, we have an overlapping very close regarding both PageRank and CheiRank. The highest value for $j=100$ is for $K_{nowc}$ vs. $K_{wc}$ with $\eta_{N} = 0.75$. \textbf{\textit{wcpv}} method gives us more differences in the top 100 with respectively 53 $\%$ and 6 $\%$ of similarity for PageRank and CheiRank algorithm. Exact overlap $\eta_{O}$ is very low with $5 \%$ and $2 \%$ regarding $K_{wcpv}$ and $K_{wc}$ with $K_{nowc}$ for $j = 100$. Exact similarity regarding CheiRank is $0.0$ for \textbf{\textit{wc}}, \textbf{\textit{wcpv}} and \textbf{\textit{nowc}} methods. The left panel shows us how similar $K_{wc}$ (Resp. ${K^{*}}_{wc}$) and $K_{wcpv}$ (Resp. ${K^{*}}_{wcpv}$) are, with clickstreams and pageviews statistical rankings. Highest measures are for CheiRanks, \textbf{\textit{wcpv}} method has the highest overlap with \textit{vR} ($0.5$ and $0.24$) for both PageRank and CheiRank.

The overlap measures show us that when we take into account social impact in PageRank and CheiRank algorithm, we have a drastic change in the final ranking. This change is mainly due to pageviews information. We are interested in \textbf{\textit{wcpv}} method, because it brings new elements to both PageRank and CheiRank.

\begin{figure}
\includegraphics[width=0.8\textwidth]{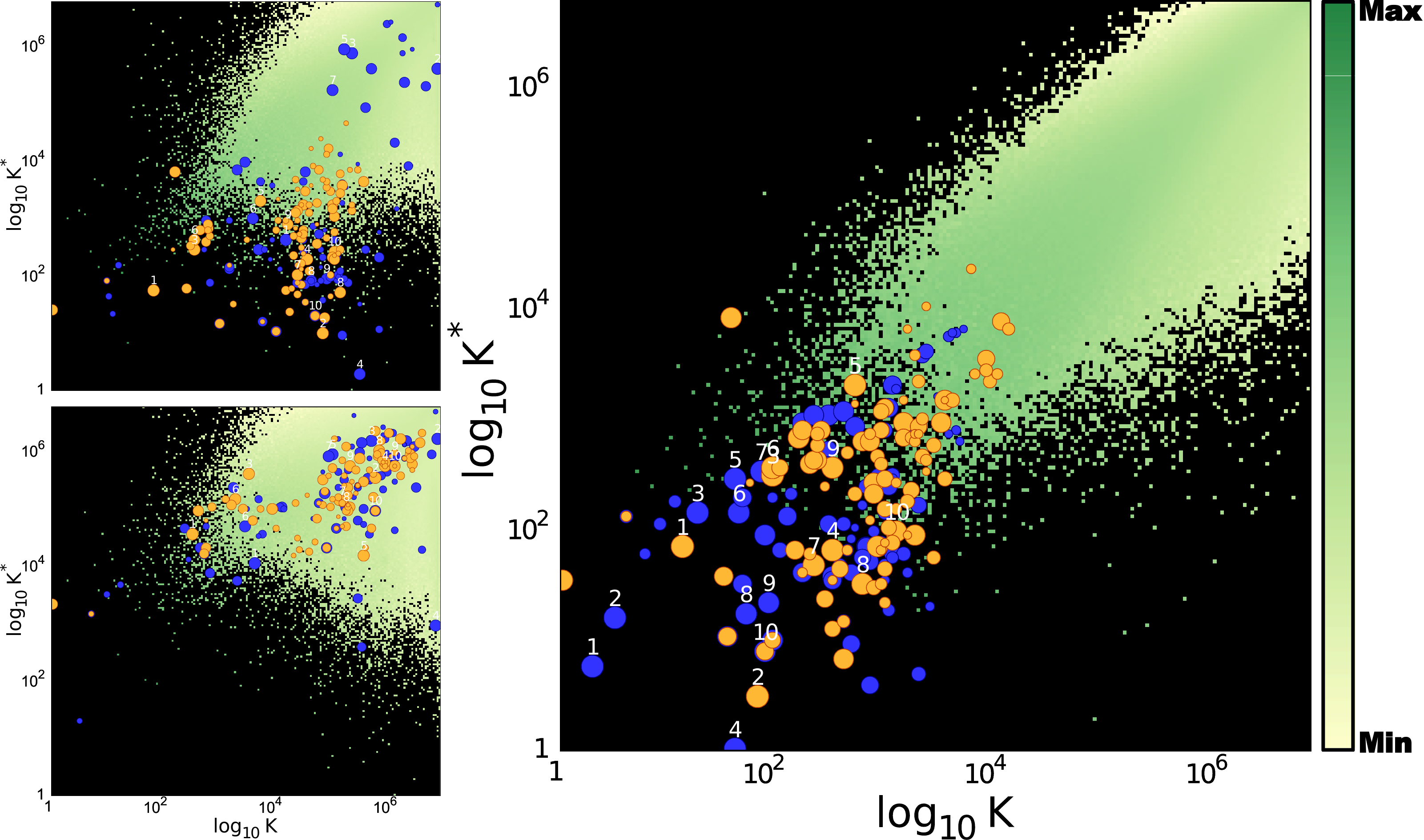}
\caption{Density distribution of Wikipedia articles $W(K,K^{*})=dn/dKdK^{*}$ in case of \textbf{\textit{nowc}} (bottom left), {\textbf{\textit{wc}}} (top left) and \textbf{\textit{wcpv}} (right) models. We have divided $K,K^{*}$ plane into $200\times 200$ decimal logarithmic cells. For each cell of area $dKdK^{*}$ we compute the corresponding articles density. We colored boxes using a decimal logscale. Blue and orange circles respectively represent the top 100 articles from \textit{vR} and \textit{cR}. Top 10 of both \textit{vR} and \textit{cR} are labeled. Circle radius is decreasing with the rank.}
\label{fig2}
\end{figure}

Fig.~ \ref{fig2} shows the distribution of articles in the $(K, K^{*})$ plane. In case of \textbf{\textit{nowc}}, articles with good PageRank have bad CheiRank and reversely. The \textbf{\textit{nowc}} method doesn't represent the social interest and is robust with time. In case of both \textbf{\textit{wc}} and \textbf{\textit{wcpv}} methods, the articles in $(K, K^{*})$ plane related to top \textit{cR} and \textit{vR} have a lower $K$ and $K^{*}$ value. As we can see with bottom panel, {\textbf{\textit{wcpv}}} method brings the top 100 of \textit{vR} and \textit{cR} at the left bottom corner of ($K,K^{*}$) plane. Also the articles are much more along the diagonal. We think that using a nontrivial matrix teleportation, \textbf{\textit{wcpv}} method tends to give a PageRank and a CheiRank for an article that are much more similar compared to previous methods.

2Drank algorithm, based on both $K$ and $K^{*}$, gives a higher rank to an article that is central in term of incoming and outgoing links. In case of \textbf{\textit{wcpv}} method applied on English edition of Wikipedia, the top 10 contains only one sovereign country, which is "United States". The first element is ''Wikipedia'', which is expected, but also missing from \textbf{\textit{nowc}} 2Drank's top 10 ($478^{th}$). The top 10 of \textbf{\textit{wcpv}} 2Drank is far from the \textbf{\textit{nowc}} and \textbf{\textit{wc}} ones but closer to top \textit{vR}.
Articles of social interests present in this top 10 are ''September 11 attacks'', ''Donald Trump", ''Greta Thumberg'' and ''It Chapter Two''.

\textbf{SQL dumps}. In case of the network based on SQL data, we see a difference when we keep information related to the ''Main Page''. For English edition SQL, this article isn't a dangling node anymore. Wikipedia regularly suggests to users a set of articles in this main page. By removing ''Main Page'' related information, some interesting high PageRanked articles are missing: ''Brexit'', ''Impeachment inquiry against Donald Trump'', ''2019 Southeast Asian haze'' for \textbf{\textit{wcpv}} PageRank and 'English Wikipedia'' and ''QR code'' for \textbf{\textit{wcpv}} CheiRank. Obviously Main Page is the top 1 for \textbf{\textit{wcpv}} PageRank and CheiRank. 

In case of applying our method to SQL dumps, keeping ''Main Page'' leads to new information. Larger tables of top 100 articles are available at \cite{urlSM}.

\subsection{Multilingual network}
A simple way to build a multilingual Wikipedia network is to aggregate networks corresponding to each considered edition (11 languages). \textbf{\textit{wcpv}} method weights the link $A\rightarrow B$ with clicks summed over all considered editions.

Here, we show results and discuss the case of both XML and SQL based multilingual Wikipedia network considering 11 languages: Chinese, English, French, German, Italian, Japanese, Spanish, Persian, Polish, Portuguese, Russian.

Regarding table~\ref{tab:top20prjoinedwiki19xml-sql}, we see that the \textbf{\textit{wcpv}} PageRank's top 10 presents more differences for SQL than for XML version of the network. In context of SQL, the top 10 {\textbf{\textit{wcpv}}} PageRank is very different from both \textit{vR} and \textit{cR}. Moreover, top 10 related to \textbf{wcpv} shows different countries, "United Kingdom" and "France" for XML and "United States of America" and "Japan" for SQL.
As for the case of only English, $K^{*}_{wcpv}$ is very far from $K^{*}_{nowc}$ as we can see with table~ \ref{tab:top20crjoinedwiki19xml-sql}, whether it's XML or SQL. We also have fewer articles related to lists in case of the multilingual Wikipedia network. There are 2 articles related to social trend in common : "It: Chapter Two" and "Once Upon a Time in Hollywood".

The top 10 Wikipedia articles \textbf{\textit{wcpv}} using 2Drank for XML multilingual network is much closer to \textit{vR} compared to other rankings. SQL dumps leads to a top 10 \textbf{\textit{wcpv}} 2DRank far from \textit{vR}. Top 10 \textbf{\textit{wcpv}} 2DRank elements with rank in $[2,4]$ are very unexpected, their corresponding rank in other lists are at least equal to $5287$ (\textit{cR}) and at last $605770$ (\textit{vR}). These articles are of social interests ''Queen of the South", Alice Braga'', ''La Reina del Sur''. Note that Alice Braga is a main character of this TV series.
In case of SQL \textbf{\textit{wcpv}} top 10 2DRank, we found 4 sovereign countries ''United States'', ''Japan'', ''Italy'' and ''Russia''which are not present in \textbf{\textit{nowc}} 2DRank top 10.

\begin{table}[]
\caption{First 10 articles obtained by PageRank with \textbf{\textit{wcpv}} model from the Multilingual Wikipedia network obtained with XML and SQL dump. Source: own calculations in October 2019}
\centering
\resizebox{\columnwidth}{!}{
\begin{scriptsize}
\begin{tabular}{|c|p{5cm}|c|c|c|c|c|}
\hline
\multicolumn{2}{|c|}{Name}&$K_{wcpv}$&$K_{wc}$&$K_{nowc}$&$K_{cR}$&$K_{vR}$\\
\hline
\multirow{10}{*}{\rotatebox{90}{\usebox0}}
& continent&1&1&1&4&25\\
\cline{2-7}
& United Kingdom&2&6&5&27&49\\
\cline{2-7}
& endemic to&3&3&4&81&43\\
\cline{2-7}
& France&4&2&2&161&23\\
\cline{2-7}
& Wikipedia&5&5606&2680&12647&1\\
\cline{2-7}
& English&6&13&3&865&203\\
\cline{2-7}
& World War II&7&11&6&73&39\\
\cline{2-7}
& People's Republic of China&8&9&10&77&50\\
\cline{2-7}
& list of Queen of the South episodes&9&9296110&9299151&8992585&2\\
\cline{2-7}
& headquarters location&10&5&8&112&70\\
\cline{2-7}
\hline
\multirow{10}{*}{\rotatebox{90}{\usebox1}}
& List of Queen of the South episodes &1&152700&9402560&1&1\\
\cline{2-7}
& International Standard Book Number&2&2&1&12422&6820\\
\cline{2-7}
& United States of America&3&4&5&3&596\\
\cline{2-7}
&  Queen of the South &4&128752 & 782421&33017&32562\\
\cline{2-7}
&  Geographic coordinate system &5&1&4&27224&4161\\
\cline{2-7}
 & Wikidata&6&3&2&166656&578987\\
\cline{2-7}
 & Virtual International Authority File&7&5&6&215168&77748\\
\cline{2-7}
 & English&8&6&3&571&2227\\
\cline{2-7}
 & Library of Congress Control Number&9&8&7&171325&51054\\
\cline{2-7}
 & Japan &10&26&19&175&1992\\
\cline{2-7}
\hline
\end{tabular}
\end{scriptsize}
}
\label{tab:top20prjoinedwiki19xml-sql}
\end{table}

\begin{table}[]
\caption{First 10 articles obtained by CheiRank with \textbf{\textit{wcpv}} model from the Multilingual Wikipedia network obtained with XML and SQL dump. Source: own calculations in October 2019}
\centering
\resizebox{\columnwidth}{!}{
\begin{scriptsize}
\begin{tabular}{|c|p{5cm}|c|c|c|c|c|}
\hline
\multicolumn{2}{|c|}{Name}&$K^{*}_{wcpv}$&$K^{*}_{wc}$&$K^{*}_{nowc}$&$K_{cR}$&$K_{vR}$\\
\hline
\multirow{10}{*}{\rotatebox{90}{\usebox0}}
& Deaths in 2019&1&1&571&1029&5\\
\cline{2-7}
& Lists of deaths by year&2&2&12&88018&20254\\
\cline{2-7}
& It: Chapter Two&3&6&318108&6&15\\
\cline{2-7}
& It&4&12&87569&18&30\\
\cline{2-7}
& Once Upon a Time in Hollywood&5&9&78393&15&37\\
\cline{2-7}
& 2019 FIBA Basketball World Cup&6&3&22945&47&11\\
\cline{2-7}
& Joker&7&16&247252&35&12\\
\cline{2-7}
& Greta Thunberg&8&23&94431&869&7\\
\cline{2-7}
& September 11 attacks&9&10&10665&21&13\\
\cline{2-7}
& 2019 in film&10&15&249&750&83\\
\cline{2-7}
\hline
\multirow{10}{*}{\rotatebox{90}{\usebox1}}
& Deaths in 2019&1&1&388&284&914\\
\cline{2-7}
& Queen of the South&2&7495&219841&33017&32562\\
\cline{2-7}
& List of Queen of the South episodes&3&433980&5166620&1&1\\
\cline{2-7}
& List of Bollywood films of 2019&4&17&12084&1640&4\\
\cline{2-7}
& Llists of deaths by year&5&2&1231&64356&1070417\\
\cline{2-7}
& 2019&6&4&2195&821&4652\\
\cline{2-7}
& It: Chapter Two&7&5&388604&7&360\\
\cline{2-7}
& Once Upon a Time in Hollywood&8&7&78976&17&66\\
\cline{2-7}
& 2019 in film&9&13&2578&262&592877\\
\cline{2-7}
& September 11 attacks&10&9&12762&23&320\\
\cline{2-7}
\hline
\end{tabular}
\end{scriptsize}
}
\label{tab:top20crjoinedwiki19xml-sql}
\end{table}

\begin{figure}
\includegraphics[width=0.7\textwidth]{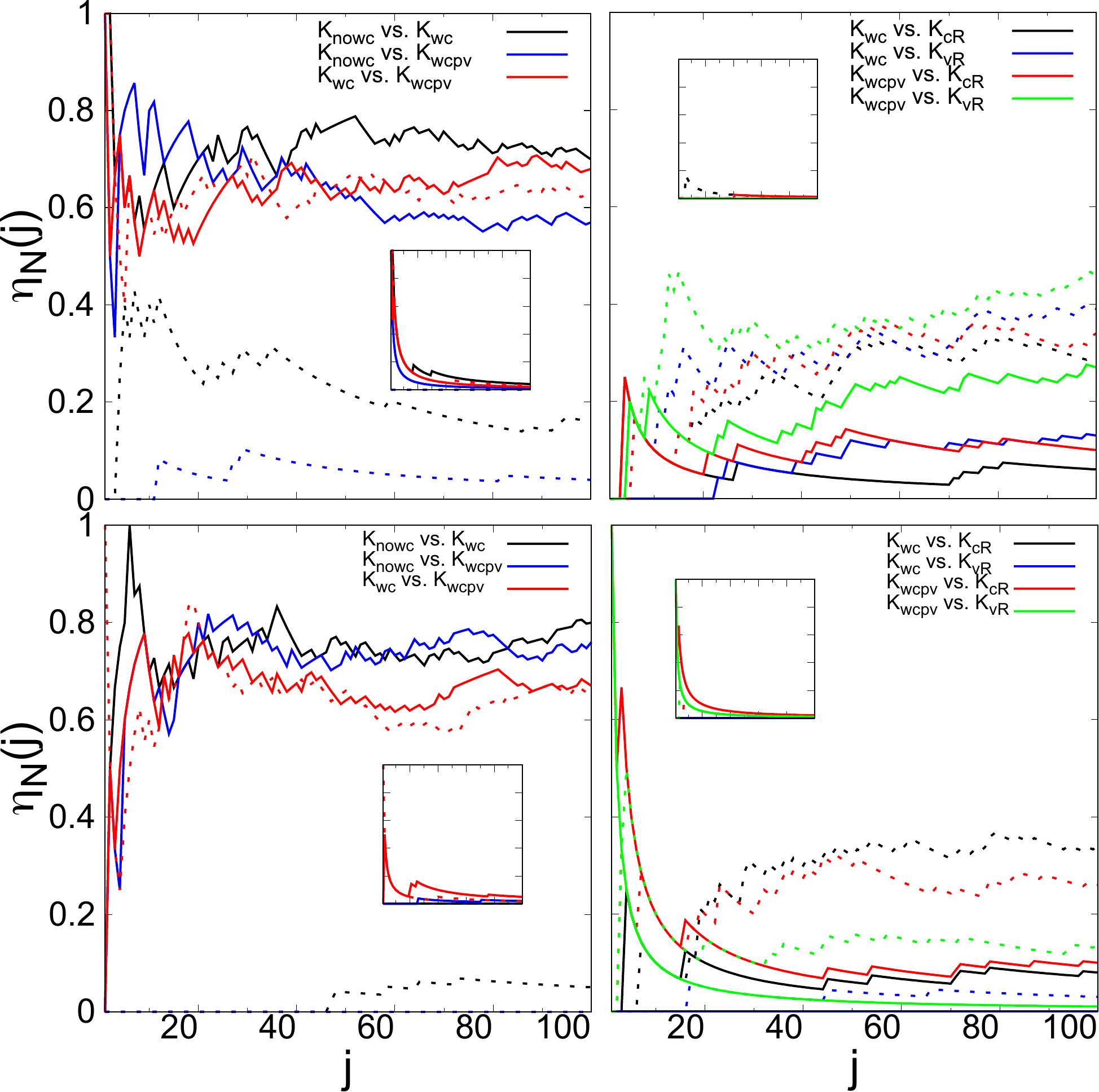}
\caption{Overlap $\eta_{N}$ versus rank $j$ for doublet of ranking lists computed from \textbf{\textit{wc}}, \textbf{\textit{wcpv}} and \textbf{\textit{nowc}} models (left panels),  from {\textbf{\textit{wc}}}, {\textbf{\textit{wcpv}}}, \textit{cR} and \textit{vR} (right panels). Inset plots correspond to exact overlap $\eta_{O}$. Top row is for multilingual Wikipedia network built with XML dumps and bottom row is for network built with SQL dumps.} \label{fig3}
\end{figure}

Regarding fig.~\ref{fig3}, as for the English version of Wikipedia network, we see that CheiRanks' overlapping with other CheiRanks and with both \textit{cR} and \textit{vR} are lower than for PageRanks. In case of multilingual version, the exact overlap $\eta_{O}$ is higher but still low compared to $\eta_{N}$ values.
Regarding both SQL and XML dumps, \textbf{\textit{wc}} is the most similar to \textbf{\textit{nowc}} with respectively $\eta_{N} = 0.8$ and $0.7$. While PageRanks are more similar and CheiRanks related overlaps are almost $0$ for SQL, XML version leads to more different tops 100 for PageRank. \textit{cR} and \textit{vR} have the same similarities with $K_{wc}$ (resp. $K^{*}_{wc}$) and $K_{wcpv}$ (resp. $K^{*}_{wcpv}$) for both XML and SQL. In case of XML, $K_{wcpv}$ and \textbf{vR} are much closer than $K_{wc}$ and \textbf{vR}.

\begin{figure}
\includegraphics[width=0.9\textwidth]{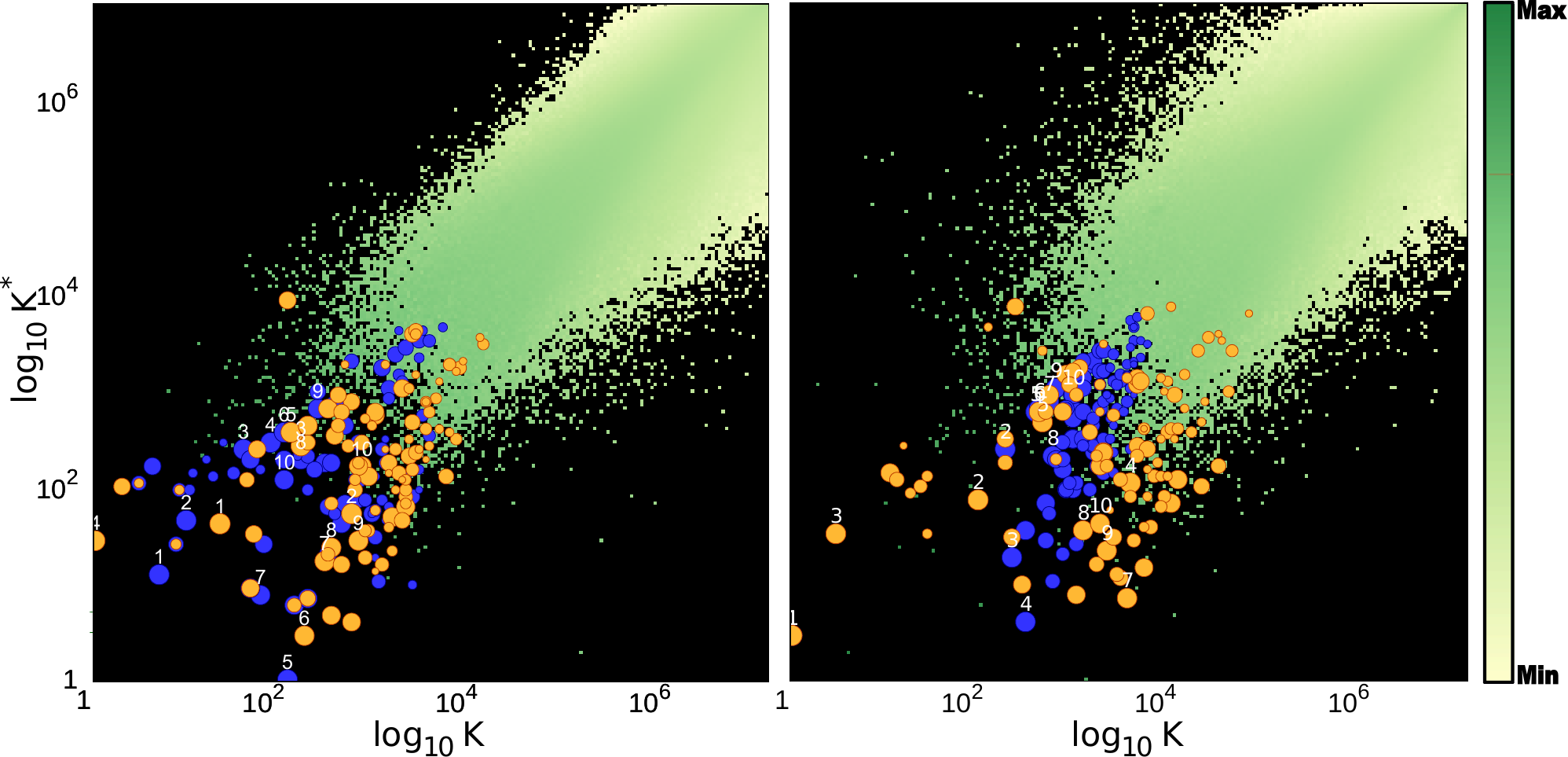}
\caption{Density distribution of Wikipedia articles in case of {\textbf{\textit{wcpv}}} XML dumps (left) and SQL dumps (right). The same color code as in fig.~\ref{fig3}.} \label{fig4}
\end{figure}

We present in fig.~\ref{fig4} the articles' distribution in $(K, K^{*})$ plane for {\textbf{\textit{wcpv}}} in case of both XML and SQL multilingual Wikipedia network. The articles are still organized along the diagonal line and the top 100 from \textit{cR} and \textit{vR} are gathered in the bottom left corner of the plots. Unlike SQL dumps, in case of XML dumps, the green circles corresponding to top 100 \textit{vR} have better PageRank than articles from top 100 \textit{cR}. The top 100 \textit{vR} articles are more scattered in case of XML than SQL dumps.

In case of a multilingual Wikipedia network, \textbf{\textit{wcpv}} method brings articles of social interest to the top of PageRank and CheiRank. Moreover, by changing the dump from SQL to XML, articles from both top \textit{vR} and top \textit{cR} have better PageRanks.

\section{Conclusion and future works}
Wikipedia gives us plenty of free information related to a large spectrum of knowledge. Articles in this free encyclopedia are edited, checked and corrected by various users (even anonymous).A standard use of PageRank and other related ranking algorithms give a time robust ranking of articles whereas clickstream and pageviews based ranking reflects statistical social trends. In this study, we present an altered version of PageRank, CheiRank and 2DRank by using both clickstream and pageviews data together with connections between Wikipedia articles. ''WikiClick Plus View'' (\textbf{\textit{wcpv}}) model of the Wikipedia network gives different rankings of Wikipedia articles. With \textbf{\textit{wcpv}} we measured the centrality of articles in Wikipedia network regarding the actual social interest. We showed that the two type of Wikipedia dumps (XML and SQL) may give different results in final rankings. \textbf{\textit{wcpv}} model gives the top PageRank articles related to actual social interest and the top CheiRank articles that are not related to lists (as they usually are) but rather related to entry point of interest. Instead of roughly aggregating individual language edition based Wikipedia networks to build a multilingual network, the use of \textbf{\textit{wcpv}} method permits us to define a more realistic linkage between articles, by using clickstream data as weight and pageviews as teleportation matrix. Merging both the ephemeral aspect of social trends with time robustness of links based on historical truth, we think that this method can be used for further interesting results.

In future works, we plan to provide a deeper analysis on advantages that can give proposed novel versions of PageRank, CheiRank and 2DRank using social impact in different fields. In this work, we only used data from October of 2019. In our next works, we plan to investigate differences between results of the measures from various time periods of the content of the Wikipedia articles, pageviews and clickstream data. Based on such time dependent data, we would be able to find out the correlation between social impact and evolution of the articles on selected language versions and also on multilingual level. Such data can also be useful to analyse how the indirect social flow between articles lead to the creation of new links. Another interesting direction of the research would be to find out how the proposed measures influences quality of the content in Wikipedia. Using these measures as additional predicting variables could also improve existing prediction models of stock market moves and performances (including price of cryptocurrencies), success of the products or demand for services, as well as electoral predictions and forecasting epidemics in the specified territory. 

Wikipedia is one of the representatives of wiki services. Therefore methods proposed in the paper can be also valuable for any knowledge base created using MediaWiki open source software, including corporate ones. These knowledge bases can contain information about customers, products and other business oriented content. Therefore our method can provide new information to companies allowing them to understand social trend's evolution and help to improve products recommendations for customers, as well as improve existing prediction models related to stock market moves, demand for services, elections results and others.

{\textit{Acknowledgments:}}
We thank D. Shepelyansky and J. Lages for useful discussions. This work was supported by the French “Investissements d’Avenir” program, project ISITE-BFC (contract ANR-15-IDEX-0003) and by the Bourgogne Franche-Comté Region 2017-2020 APEX project (conventions 2017Y-06426, 2017Y-06413, 2017Y-07534; see \url{http://perso.utinam.cnrs.fr/~lages/apex/}).
%
%
\bibliographystyle{splncs04}
\bibliography{mybibliography}
%
%
\end{document}